\documentclass[sigconf]{acmart}

\usepackage{algorithm}
\usepackage{algorithmic}
\usepackage{subcaption}
\usepackage{bbold}

\newtheorem{proposition}{Proposition}
\AtBeginDocument{%
  \providecommand\BibTeX{{%
    \normalfont B\kern-0.5em{\scshape i\kern-0.25em b}\kern-0.8em\TeX}}}

\copyrightyear{2022} 
\acmYear{2022} 
\setcopyright{acmlicensed}\acmConference[WWW '22]{Proceedings of the ACM Web Conference 2022}{April 25--29, 2022}{Virtual Event, Lyon, France}
\acmBooktitle{Proceedings of the ACM Web Conference 2022 (WWW '22), April 25--29, 2022, Virtual Event, Lyon, France}
\acmPrice{15.00}
\acmDOI{10.1145/3485447.3512097}
\acmISBN{978-1-4503-9096-5/22/04}

\begin{document}

\title{Adaptive Experimentation with Delayed Binary Feedback}


\author{Zenan Wang}
\orcid{1234-5678-9012}
\affiliation{%
  \institution{JD.com}
  \city{Mountain View}
  \state{CA}
  \country{USA}
}
\email{wangzenan5@jd.com}

\author{Carlos Carrion}
\affiliation{%
  \institution{JD.com}
  \city{Mountain View}
  \state{CA}
  \country{USA}
}
\email{carlos.carrion@jd.com}

\author{Xiliang Lin}
\affiliation{%
  \institution{JD.com}
  \city{Mountain View}
  \state{CA}
  \country{USA}
}
\email{xiliang.lin@jd.com}

\author{Fuhua Ji}
\affiliation{%
  \institution{JD.com}
  \city{Beijing}
  \country{China}
}
\email{jifuhua@jd.com}

\author{Yongjun Bao}
\affiliation{%
  \institution{JD.com}
  \city{Beijing}
  \country{China}
}
\email{baoyongjun@jd.com}

\author{Weipeng Yan}
\affiliation{%
  \institution{JD.com}
  \city{Beijing}
  \country{China}
}
\email{paul.yan@jd.com}
\renewcommand{\shortauthors}{Wang et al.}

\begin{abstract}
Conducting experiments with objectives that take significant delays to materialize (e.g. conversions, add-to-cart events, etc.) is challenging. Although the classical ``split sample testing" is still valid for the delayed feedback, the experiment will take longer to complete, which also means spending more resources on worse-performing strategies due to their fixed allocation schedules. 
Alternatively, adaptive approaches such as ``multi-armed bandits" are able to effectively reduce the cost of experimentation. But these methods generally cannot handle delayed objectives directly out of the box. This paper presents an adaptive experimentation solution tailored for delayed binary feedback objectives by estimating the real underlying objectives before they materialize and dynamically allocating variants based on the estimates. Experiments show that the proposed method is more efficient for delayed feedback compared to various other approaches and is robust in different settings. In addition, we describe an experimentation product powered by this algorithm. This product is currently deployed in the online experimentation platform of JD.com, a large e-commerce company and a publisher of digital ads.
\end{abstract}




\maketitle

\section{Introduction}
Experimentation has been widely used in the tech industry and especially for content optimization in websites and online advertising. A typical experiment system will assign users or requests to different groups and display one variant of contents among several possibilities accordingly. Then users' interactions with the content such as clicks and purchases etc. are collected to construct metrics like click-through rate (CTR), conversion rate (CVR),  user return rate, dwelling time, etc. for analyzing user engagement \cite{lehmann2012models}. One key aspect of the system that does not receive a lot of attention is that there could be significant delays between a user's visit to the page and their actions. A click may be instantaneous, but a purchase could take hours or even days for a user to complete. Using objectives with delays could introduce problems to the experiment.

The traditional process of assigning a fixed portion of users to competing alternatives is also known as online A/B/n
testing and is readily available in major online experimentation platforms \cite{kohavi2020}. The biggest advantage of A/B/n testing is that it's easy to implement and can easily support a variety of metrics of interest, including objectives with delays. 
However, using a delayed objective in an A/B/n test means it takes longer to finish the experiment than it otherwise would, which in turn exacerbates two common problems that A/B/n testing is criticized for. First, a big complaint of the A/B/n testing is that it could incur sizeable experimentation costs. Because all the alternatives in the A/B/n tests are guaranteed a fixed portion of sample size, a treatment may be allocated to a significant portion of the users even if it turns out to be a ``bad'' treatment and hurtful to the user experience. A longer experiment means even larger experimentation costs. Second, A/B/n testing is prone to peeking, which inflates the Type-I error. Because the A/B/n tests are designed to be analyzed only when the experiments end, peeking results and making decisions before the end of an experiment could lead to erroneous conclusions. Having to run a longer experiment for delayed objectives creates more opportunities for mistakes.
Although there are advanced methods to address the peeking issue, such as sequential tests \cite{johari2015always},  
as far as we know, the prominent methods today may not work easily with delayed feedback. 

In recent years, adaptive tests have gained traction. Especially, ``multi-armed bandits" test designs \cite{scott2010,scott2015multi, geng2021comparison} increasingly becomes an alternative to the A/B/n testing when experimenters are only concerned with one primary metric. These bandit tests have the key advantage of reducing the opportunity costs from the experimentation, allocating traffic that would be diverted to inferior variants to variants with more potential gains, as \cite{scott2015multi} points out. However, widely-used ``multi-armed bandits" test designs require the metric or objective of interest to provide feedback in a reasonable time frame in order to update the assignment rule to the variants. Consequently, adaptive tests have found the most success with metrics with near-instantaneous feedback, especially CTR. 

Developing bandit algorithms for delayed feedback has become a hot topic recently \cite{lattimore2020}, for which we give an
overview in Section \ref{sec:literature}. But there are very few practical solutions that are directly applicable to our use case, optimizing for CVR. Aside from the fact that conversions are often delayed, another complexity for the CVR objective is that we would never observe negative feedback. If we have not observed a conversion from one particular user, it's because either she will convert in the future or she would never convert to begin with. And it is impossible to distinguish between these two possibilities. Metrics with such a property are common. For example, computing user return rates also face a similar issue: a user may return at some time in the future or she may never return, but both cases are observably the same \cite{dupret2013absence}.

In this paper, we present a \emph{Thompson Sampling} (TS) algorithm to extend the ``multi-armed bandits" test designs to binary metrics with significantly delayed feedback. In our practical application discussed in detail subsequently, we focus on conversion rate (CVR) as the key metric. Nevertheless, the underlying ideas of the proposed TS algorithm are readily applicable to other binary metrics with delayed feedback and can be extended for delayed continuous metrics. 

We contribute to the literature by formulating a common real-world problem and proposing a coherent and cogent solution that is practically appealing and easy to implement. 
Our proposed algorithm leverages a hybrid model within a Thompson Sampling Bandit framework \cite{scott2010, lattimore2020}. The key features of our proposed algorithm are 
\begin{enumerate}
    \item Modeling the objective using two latent variables, the eventual conversion, and the delay;
    \item Computing the delay-corrected feedback during the experiment using Expectation-Maximization method;
    \item Selecting the optimal creatives based on the delay-corrected feedback.
\end{enumerate}
We use simulations to benchmark our proposed algorithm against other approaches, and also present a real-world example of advertisers using our proposed algorithm after it is deployed. Our solution is deployed in the online experimentation platform of JD.com, a large e-commerce company and a publisher of digital ads, and it allows advertisers to optimize their ads creatives exposed to on average tens of millions of requests per day.

\section{Related Works}\label{sec:literature}
This paper belongs to the fast-growing literature of using bandit algorithms for web-content optimization (e.g. personalized news recommendation, personalized creatives for ads, etc.) \cite{agarwal2009explore, li2010contextual,schwartz2017customer,  scott2015multi, geng2020}. 
However, unlike this paper, almost all the applied bandit papers are focusing on optimizing instantaneous metrics, and more specifically CTR (see \cite{chapelle2011empirical,agarwal2009explore, geng2020, geng2021comparison} for example), because a key assumption behind their algorithm and analyses is that the reward needs to be immediately available after an action is taken. 
For advertisers and decision-makers, CVR, sales, and other delayed metrics are often more important than CTR because those are more directly related to the business Key Performance Indicators (KPIs). CTR is used as a proxy for its convenience, but may not lead to the optimum in the desired KPIs. In \citet{schwartz2017customer}'s analysis, they found that customer acquisition would drop 10\% if the firm were to optimize CTR instead of conversion directly.

In the broader online learning algorithms literature, there are a handful of research projects extending bandit algorithms to delayed feedback \cite{joulani2013online, vernade2017stochastic,pike2018bandits,  liu2019multi, vernade2020linear}. These research projects seek to address delayed feedback issues under different settings and mostly focus on theoretical analysis of the algorithms through the derivation of complexity bounds for regret. In the influential empirical study of Thompson Sampling \cite{chapelle2011empirical}, the authors discuss the impact of the delays on the TS algorithm, but they only consider the fixed and non-random delays. A more general problem of learning with delays is discussed in \cite{joulani2013online}, but the proposed modifications only apply to the delays with known parameters. In the more recent work \citep{zhou2019learning}, the authors consider stochastic delays which are more reasonable for practical applications. But unlike in our setting, all the delays are assumed to be observed eventually, which is not applicable for CVR because non-converted clicks are never observed.

The closest works to ours in terms of the problem settings are \cite{chapelle2014}, \cite{vernade2017stochastic} and \cite{vernade2020linear}, where the feedback are not only stochastically delayed, but also can be missing entirely. \citet{chapelle2014} proposes to treat the conversion rate estimation as an offline supervised learning problem, and set it up as a variation of the statistical censoring problem. Similar to our paper, \citet{vernade2017stochastic} tries an online approach, but they focus more on the theoretical properties and assume the delay distribution is known, which is not very applicable in practice. The authors' recent follow-up paper \cite{vernade2020linear} relaxes the assumption to allow for unknown delay distribution, but introduces a hyperparameter $m$, which is essentially a timeout limit. If feedback has not been received within $m$ rounds, their algorithm
will label it as a non-conversion. It is an interesting approach but has some limitations for practitioners to use. First, it's not clear how to choose a proper $m$. Second, the estimated CVR is biased, and more likely to underestimate the conversion.

\section{Problem Setup and Challenges}\label{sec:setup}
In the example used throughout this paper, our goal is to find the creative with the best CVR among a set of competing alternatives. Following the standards of the online advertising industry, we use the post-click attribution model, which means that CVR is defined as the percentage of the converted clicks among all the clicks. And a click is considered converted if a purchase occurred at some time after the click.

One unique aspect of the CVR (and other delayed binary feedback) problem is that the positive feedbacks take time to be observed while the negative feedbacks are never observed. Therefore, we use a hybrid model with two latent random variables to capture this dynamic. Formally, for each click $i$ in the experimental group $k$, the outcome of the click is characterized by the following two latent variables:
    \begin{itemize}
        \item $C_{ik}\in \{0,1\}$, whether the click $i$ in group $k$ will eventually lead to a conversion \footnote{When a click leads to multiple orders, we still consider it to be one conversion and use the earliest order to compute the delay.};
        \item $D_{ik}\in \mathbb{R}^+$, the delay between the time of click $i$ and the time its conversion if $C_{ik}=1$.  $D_{ik}$ is not defined if $C_{ik}=0$.
    \end{itemize}  

We are interested in estimating the conversion rate for each treatment group $k$, which is defined as $\theta_{k}\equiv E[C_{ik}]$.

For the simplicity of the notation, let us only consider one treatment group for the ensuing discussion and omit the group subscript. We will bring back the group subscript in Section \ref{subsec:bandit_integration}.

\subsection{Challenges}
A common practice to measure CVR at any given time $t$ in the online advertising industry is to compute the \textit{naive CVR}, i.e. $\widetilde{\theta_t}\equiv\frac{N^{convert}_{t}}{N_t}$,
where $N_t$ and $N^{convert}_{t}$ respectively represent the total number of clicks and conversions up until time $t$. Throughout this paper, we will use lowercase letters such as $n_t$ to represent contemporaneous counts at time $t$, and uppercase letters such as $N_t$ to represent the cumulative counts up to time $t$.

\begin{figure}
    \centering
\includegraphics[width=\linewidth]{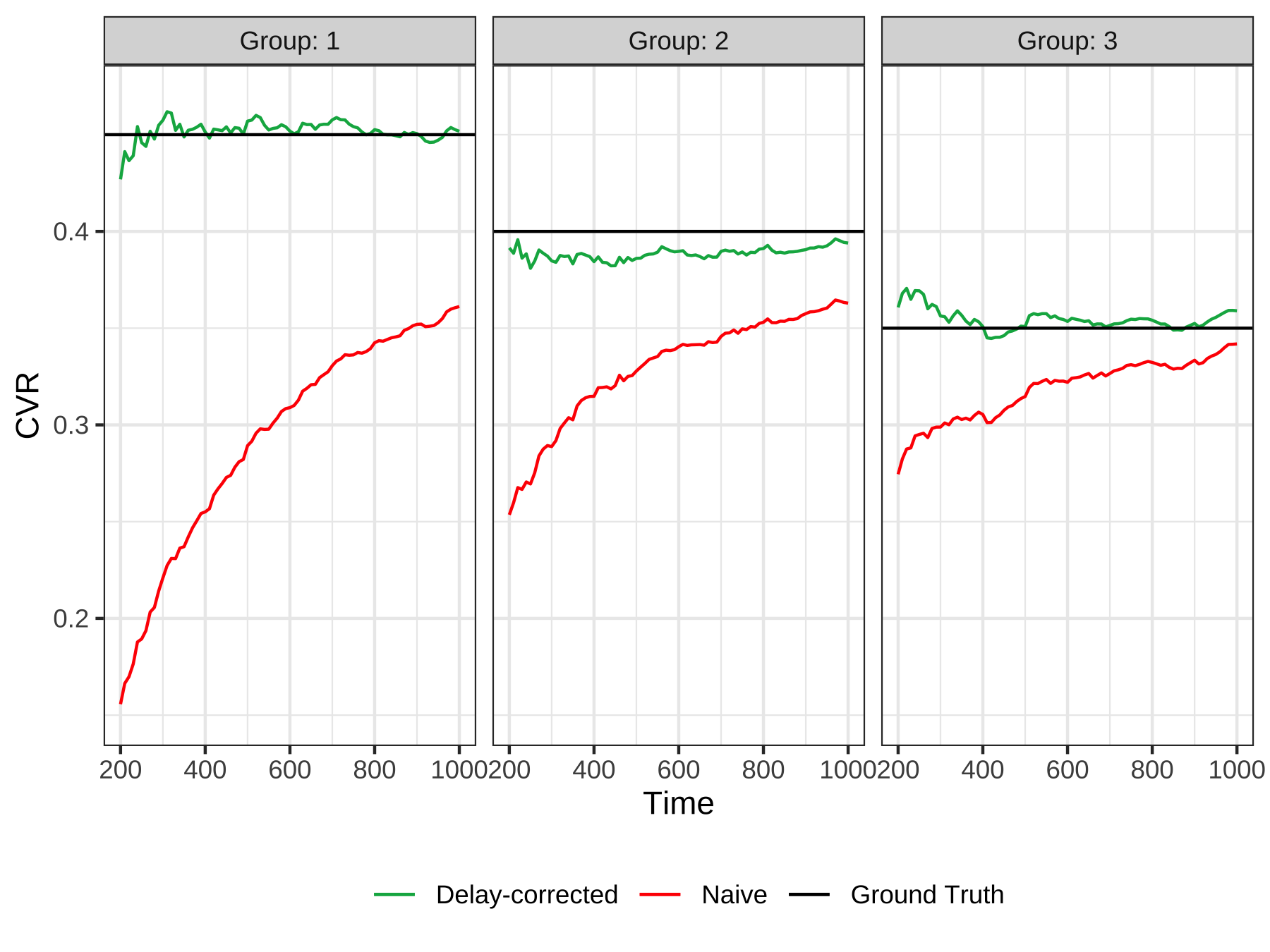}
    \caption{The Limitation of the Naive CVR}
    \label{fig:cvr_comparison}
\end{figure}

Using the latent variables defined above, we can rewrite the \textit{naive CVR} as\footnote{Here we abuse the notation and use $i^s$ to index the clicks that occurred at time $s$.} 
$$\frac{1}{N_t}\sum_{s=1}^{t}\sum_{i^s=1}^{n_s} C_{i^s} \mathbb{1}\left\{D_{i^s} \leq t-s\right\},$$ 
and thus it is trivial to show that $\widetilde{\theta_t}$ is an unbiased estimator of $\theta$ only when there is no conversion delay; whereas when there is any delay, it systematically underestimates the $\theta$. Therefore, the \textit{naive CVR} is not suitable to be used with the bandit algorithm as an outcome metric if the real conversion is delayed. As shown by the red line in the Figure \ref{fig:cvr_comparison}, using the \textit{naive CVR} as the reward may not help identify the best alternative, when the delay distributions vary across competing treatment groups.

This problem can be addressed if the delay distribution is known. We can use the delay distribution to calculate an unbiased delay-corrected CVR estimator. 
For example, following \citet{vernade2017stochastic},
\begin{align}
    \widehat{\theta_t}=\frac{N_{t}^{convert}}{\sum_{s=1}^{t}n_sP(D \leq t-s)}\label{eqn:theta_delay_corrected}
\end{align}
The proof of unbiasedness of this estimator is presented in Appendix \ref{appendix:unbiasedness}.
The green line in Figure \ref{fig:cvr_comparison} shows that delay-corrected CVR indeed performs much better in recovering the ground truth, and thus identifying the best treatment group.

However, in practice the delay distribution is unknown. Moreover, the delay distributions could reasonably be very different across treatment groups and experiments because the treatment could leave impacts on the delays too.
As a result, we could not simply use a delay distribution estimated from historical data, but have to estimate a delay distribution for each treatment group during each experiment instead. 

During an active experiment, the delay time is right-censored at any given time, i.e. we cannot observe conversion delays longer than a threshold at any time of estimation. In the canonical survival analysis problems, all the events will eventually happen, so the right-censored portion implies the tail probability of the delay distribution \citep{prentice2011}. In contrast, in our problem, the clicks that are not yet converted (i.e. right-censored portion) may either convert in the future or not convert at all. And the composition of those two types depends on the unknown true conversion rate. Therefore, in order to accurately estimate the delay distribution, we need to know the conversion rate first. We have come full circle.

\section{Method}
\begin{figure*}
    \centering
    \includegraphics[width=0.8\textwidth]{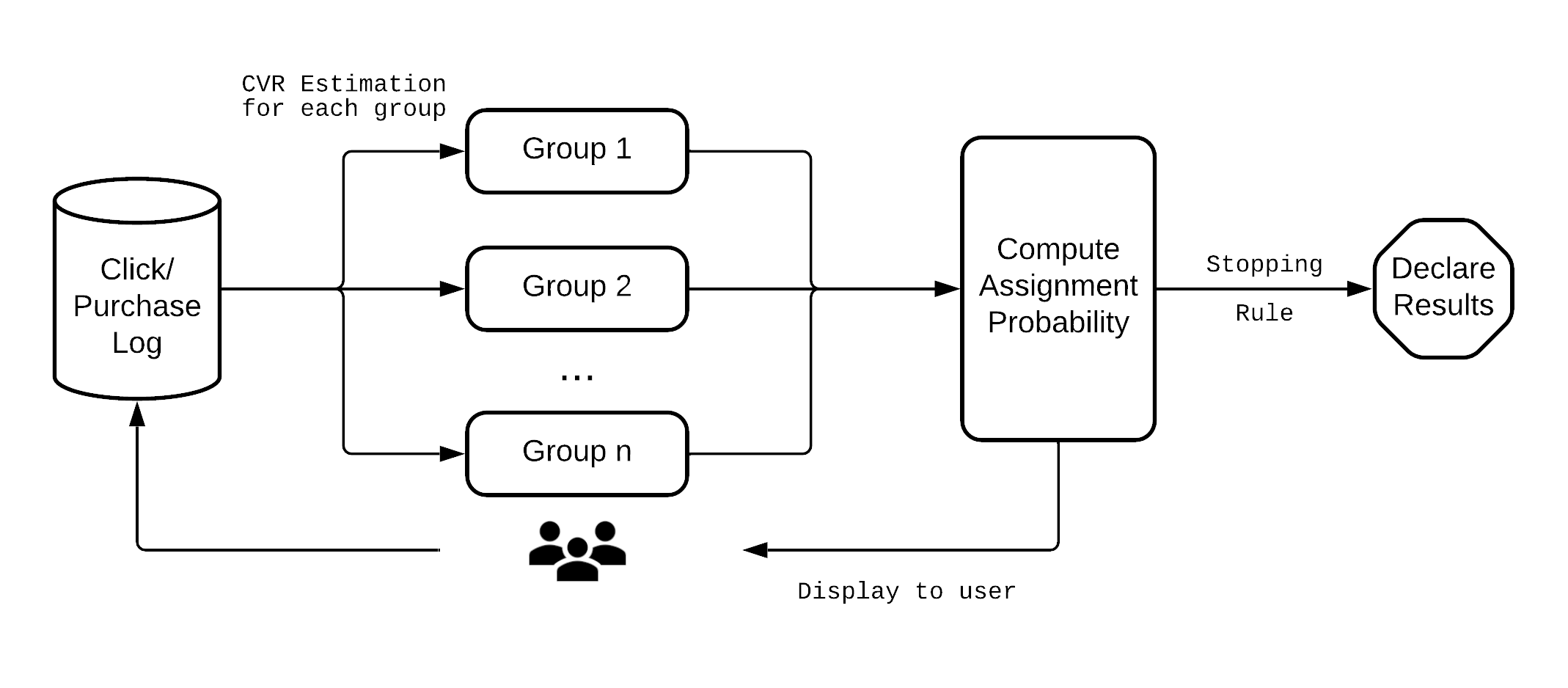}
    \caption{Method Overview}
    \label{fig:overview}
\end{figure*}

In this section, we describe the system we proposed and implemented to conduct adaptive experiments with the CVR as the target metric.
As shown in Figure \ref{fig:overview}, our system has two major components on top of the standard ad-serving infrastructure. The first component takes the click and purchase logs as inputs and estimates CVRs for each treatment group in an experiment. The second component computes the assignment probability based on all the estimated CVRs from the first component. If a stopping rule is not met, new ads will be displayed to users according to the assignment probability. Then the process repeats. Each such cycle represents a time step in our notations. It should be noted that the specific stopping criterion is independent of our proposed algorithm and should be set in accordance with the specific application. For example, an experiment can be set to stop whenever the top-performing treatment receives more than 95\% assignment probability for 24 hours.

We will describe each component in detail in the following subsections.

\subsection{CVR estimation}
In this subsection, we describe the approach to estimating CVR for each group. Because the same procedure is applied to all the treatment groups in an experiment, we will focus on one treatment group and continue omitting the group subscript for the simplicity of illustration.

As we have mentioned in Section \ref{sec:setup}, 
there are two latent variables for click $i$ in each treatment group, eventual conversion indicator $C_i$ and conversion delay $D_i$. We assume the data generating process is that, whenever a click occurs, noted as $i$,  a Bernoulli variable $C_i$ will be drawn, indicating whether this click will eventually become a conversion. Then if the click will convert, a continuous variable $D_i$ will be drawn and dictate how long it takes for the conversion to be observed.

Formally, we assume both variables are independent and identically distributed across $i$ and follow,
\begin{align*}
    C_i &\sim Bernoulli(\theta)\\
    D_i|C_i=1 &\sim \text{distribution with CDF } F(\cdot;\lambda)
\end{align*}
The $\theta$ is the unknown true CVR that we want to estimate, and $\lambda$ is a parameter that characterizes the delay distribution. We do not require the delay distribution to be any specific form except that it can be parameterized. Note that, because of the Bernoulli assumption, the above-described data generating process is only suitable for binary feedback. It's possible to extend our framework to delayed continuous feedback by choosing a different distribution for $C_i$. We discuss this possibility in Section \ref{sec:extensions}. 

Both $C$ and $D$ are not always observable at an observation time $t$. Instead, we observe the following variables:
\begin{itemize}
    \item $Y_{it}\in\{0,1\}$ indicating whether click $i$'s conversion has already occurred at $t$;
    \item $E_{it}\in \mathbb{R}^+$ is the elapsed time since the click $i$ till $t$ if $Y_{it}=0$, and the elapsed time since the click till conversion if $Y_{it}=1$, i.e.:
    \begin{align*}
    E_{it} = 
        \begin{cases}
        t - t_{i}^{click}, & \text{if } Y_{it}=0\\
        D_{i}, & \text{if } Y_{it}=1
        \end{cases}
    \end{align*}
\end{itemize}

We apply the Expectation-Maximization (EM) method to find the maximum likelihood estimates for $\theta$ and $\lambda$ \citep{hunter2004tutorial}. At any given observation time $t$, EM solves a maximum likelihood problem of the form:
$$\max_{\theta, \lambda} \sum_{i}^{N_{t}} \log\big\{ P(C_i=0, y_{it}, e_{it};\theta, \lambda )+P(C_i=1, y_{it}, e_{it};\theta, \lambda )\big\} $$

After some reformulation and applying Jensen's inequality, the above objective function is equivalent to:
\begin{align}
    \max_{\theta, \lambda} &\sum_{i}^{N_{t}} \big\{q(C_i=0)\log P(C_i=0, y_{it},e_{it};\theta, \lambda ) \nonumber\\
    &+ q(C_i=1)\log P(C_i=1, y_{it},e_{it};\theta, \lambda )\label{eqn: EM-obj}\\
    &- q(C_i=0)\log q(C_i=0)\nonumber\\
    &-q(C_i=1)\log q(C_i=1)\big\} \nonumber
\end{align}
where $q(c) = P(c|y,e;\theta,\lambda)\propto P(c,y,e;\theta,\lambda)$.

The EM method will iterate through the Expectation Step and the Maximization Step to find the solution to the above maximization problem. We detail those two steps below.
\subsubsection{Expectation Step}
For a given click and its corresponding data point  $(y_{it}, e_{it})$, we need to compute the posterior probability of the eventual conversion conditioned on the observed data:
$$w_{it} \equiv P(C_i=1|y_{it},e_{it};\theta,\lambda)$$
When $y_{it}=1$, $w_{it}$ simply equals $1$, because it is trivial that $C_{i}=1$ for certain.
When $y_{it}=0$, 
\begin{align}
    w_{it}&=Pr(C_i=1|Y_{it}=0, e_{it};\theta,\lambda)\nonumber\\
    &=\frac{Pr(Y_{it}=0, e_{it}, C_i=1;\theta,\lambda)}{Pr(Y_{it}=0, e_{it};\theta,\lambda)}\nonumber\\
    &=\frac{Pr(Y_{it}=0, e_{it}|C_i=1;\theta,\lambda)Pr(C_i=1)}{Pr(Y_{it}=0, e_{it};\theta,\lambda)}\nonumber\\
    &=\frac {(1-F(e_{it}))\theta}{1-\theta + (1-F(e_{it}))\theta}\label{eqn:e-step}
\end{align}

\subsubsection{Maximization Step}
In this step, we take the $w_{it}$ as given and 
maximize Equation \ref{eqn: EM-obj}
\begin{align*}
\max_{\theta, \lambda} \sum_{i}^{N_{t}}& (1-w_{it})\log P(C_i=0, y_{it}, e_{it};\theta, \lambda )  \\
&+ w_{it}\log P(C_i=1, y_{it}, e_{it};\theta, \lambda )    
\end{align*}
Because
\begin{align*}
P(C_i=0, y_{it}, e_{it};\theta, \lambda )
  &=\begin{cases}
    0, & \text{if } y_{it}=1\\
    1-\theta, & \text{if } y_{it}= 0 
  \end{cases}\\
 P(C_i=1, y_{it}, e_{it};\theta, \lambda )
  &=\begin{cases}
    f(e_{it}), & \text{if } y_{it}=1\\
   (1-F(e_{it}))\theta, & \text{if } y_{it}= 0 
  \end{cases}\\
\end{align*}

the objective function becomes
\begin{align}
\max_{\theta, \lambda}&\sum_{i}^{N_t} w_{it}\log\theta + (1-w_{it})\log(1-\theta)\label{eqn:m-step}\\
+&\sum_{i}^{N_t} w_{it}y_{it}\log f(e_{it}) + w_{it}(1-y_{it})\log (1-F(e_{it}))\nonumber
\end{align}

A nice result from the derivation above is that, regardless of the delay distribution $F(\cdot)$, there is always a separation between $\theta$ and $\lambda$. In other words, they can be optimized independently. This separation result comes from the fact the delay distribution is independent of the true conversion rate.

\subsubsection{Exponential Delay Distribution}
Up to this point, we have been agnostic about the distribution of the delay. Depending on the use cases and settings, one may choose different delay distributions to fit the data and our approach should work for all the parameterized delay distributions. But to give readers a more in-depth illustration of our approach work in practice, we are going to assume the delay follows an exponential distribution for the following sections.

For our use cases, we find that exponential distribution can best fit the conversion delay. \citet{chapelle2014} also reaches the same conclusion after analyzing the conversion data at Criteo. 

Plugging the \textit{probability density function} and \textit{cumulative distribution function} of 
exponential distribution into Equation \ref{eqn:m-step}, we can solve for optimal $\widehat{\lambda}_{t}^*$ analytically:

\begin{align}
 &\widehat{\lambda}^*_{t}=\frac{N_{t}^{convert}}{\sum_{i} w_{it}e_{it}}\label{eqn:lambda_mle}
\end{align}

Because of the separation, we could use the estimator described in Equation \ref{eqn:theta_delay_corrected} for $\theta$. With the exponential distribution, the estimator is:
\begin{align} 
&\widehat{\theta}^*_{t}=\frac{N_{t}^{convert}}{\sum_{s=1}^{t}n_s(1-e^{-\widehat{\lambda}_t^*(t-s)})}
\end{align}
In practice, we find that this estimator for $\theta$ is more stable than the $\theta$ estimator solved from Equation \ref{eqn:m-step}.

\subsubsection{E-M iterations}
At each time step $t$, we iterate the E-M steps for a few cycles to make sure the resulted estimates are stable. Then the final estimates are saved and used as the priors for the next time step.  
Let $L$ represent the total number of the E-M cycles. At time $t$ and cycle $l$ ($0<l\leq L$), we compute the following: 
\begin{align}
&w^{(l)}_{it}=\begin{cases}
    1, & \text{if } y_{it}=1\\
   \frac{\widehat{\theta_t^*}^{(l-1)}e^{-\widehat{\lambda_t^*}^{(l-1)}e_{it}}}{1-\widehat{\theta_t^*}^{(l-1)} + \widehat{\theta_t^*}^{(l-1)}e^{-\widehat{\lambda_t^*}^{(l-1)}e_{it}}}, & \text{if } y_{it}= 0 
\end{cases}\label{eqn:w_actual_use}\\
&\widehat{\lambda_t^*}^{(l)}=\frac{N_{t}^{convert}}{\sum_{i} w_{it}^{(l)}e_{it}}\label{eqn:lambda_actual_use}\\
&\widehat{\theta_t^*}^{(l)}=\frac{N_{t}^{convert}}{\sum_{s=1}^{t}n_s^{click}(1-e^{-\widehat{\lambda_t^*}^{(l)}(t-s)})}\label{eqn:theta_actual_use}
\end{align}
where $\widehat{\lambda_t^*}^{(0)} = \widehat{\lambda_{t-1}^*}^{(L)}$, $\widehat{\theta_t^*}^{(0)} = \widehat{\theta_{t-1}^*}^{(L)}$

\subsection{Bandit Integration}\label{subsec:bandit_integration}
After the unbiased CVRs are estimated in each treatment group for an experiment, we use a multi-armed bandit algorithm to compute the assignment probability for each group. The assignment probabilities will be used to assign requests to groups, and are updated at each time step. 

We propose to use the Thompson Sampling method with a delay-corrected sample size and a Beta-Bernoulli prior.
Specifically, we assume the eventual conversion in each treatment group follows a Bernoulli distribution with a group-specific probability $\theta_k$, consistent with what we have been assuming. And in a Bayesian framework, $\theta_k$ has a $Beta(\alpha_{kt}, \beta_{kt})$ prior at time $t$.

Before the experiment starts, at $t = 0$ we 
set diffuse priors and let $\alpha_{k0}=1, \beta_{k0}=1, \forall k \in K$. 
In the subsequent time-step $t$, we update $\alpha_{kt}$ and $\beta_{kt}$ following:
\begin{align}
    &\alpha_{kt}=1+N_{kt}^{convert}\\ &\beta_{kt}=max(1-N_{kt}^{convert}+\frac{N_{kt}^{convert}}{\theta_{kt}^*},1)
\end{align}
Then the assignment probability of a group is the posterior probability that the group offers the highest expected CVR. We compute these values using Monte Carlo simulations following the procedure outlined in \citet{scott2010}.

Algorithm \ref{alg:algorithm} presents the entire procedure of our method for exponentially distributed delays.
\begin{algorithm}[tb]
\caption{\textit{TS} to identify the group with best CVR}
\label{alg:algorithm}
\raggedright\textbf{Input}: $K$ groups\\
\textbf{Parameter}: Number of E-M cycles each step, $L$\\
\begin{algorithmic}[1] 
\STATE Let $t=0$.
\STATE $\forall k \in \mathbb{K}, \; \theta_{k0}^*\leftarrow 0.1, \; \lambda_{k0}^*\leftarrow1/105$
\STATE $\forall k \in \mathbb{K}, \; \alpha_{k0}\leftarrow 1, \; \beta_{k0}\leftarrow 1$
\STATE $\forall k \in \mathbb{K},\; p_{k0} \leftarrow 1/K$
\WHILE{NOT exit condition}
\STATE A batch of requests arrives
\FOR{request $i$}
\STATE Sample
from a multinomial distribution with $K$ groups and $p_{1t},\dots, p_{Kt}$. 
\STATE Assign $i$ to the sampled group
\ENDFOR
\STATE Collect click and conversion data
\STATE Set $t \leftarrow t+1$
\FOR{group $k$}
\STATE Load previously computed $\theta_{kt-1}^*$ and  $\lambda_{kt-1}^*$ 

\FOR{$l$ in $\{1,2,\dots,L\}$}
\STATE Update $w_{kit}^{(l)}$ for each click $i$ in $k$ as in Equation (\ref{eqn:w_actual_use})
\STATE Update $\widehat{\theta^*_{kt}}^{(l)}$ and  $\widehat{\lambda^*_{kt}}^{(l)}$ as in Equation (\ref{eqn:theta_actual_use}) and (\ref{eqn:lambda_actual_use})
\ENDFOR
\STATE $\theta_{kt}^*\leftarrow \widehat{\theta^*_{kt}}^{(L)}$,  $\lambda_{kt}^{*}\leftarrow \widehat{\lambda^*_{kt}}^{(L)}$

\STATE  Update $\alpha_{kt}=1+N_{kt}^{convert}$
\STATE Update $\beta_{kt}=max(1-N_{kt}^{convert}+\frac{N_{kt}^{convert}}{\theta_{kt}^*},1)$
\STATE
Repeatedly sample from  $Beta(\alpha_{kt},\beta_{kt})$ for all $k\in \mathbb{K}$ and  $p_{kt}$ equals the empirical proportion of Monte Carlo samples in which the draw from $k$ is maximal.
\ENDFOR
\ENDWHILE
\end{algorithmic}
\end{algorithm}

\section{Extension to Delayed Continuous Feedback}\label{sec:extensions}
The proposed algorithm described previously focuses on the case of binary delayed feedback metrics, e.g. conversion rate (CVR).
There are many important metrics such as Gross Merchandise Value that are not binary but face the same issues of delay and censoring. This algorithm can be extended to those cases of continuous metrics and even count metrics by redefining the eventual conversion $C$ variable.
The random variable $C$ could be defined as a mixed random variable with a discrete component still corresponding to the case without a response (e.g. no purchase is made), and a continuous component for the value $C > 0$ for the case with feedbacks (e.g. some amount of sales are completed).

Mathematically, $C = 0$ with a probability mass function $P(C = 0)$ and  $C \in (0, \infty)$ with a probability density function $f(c)$. Thus, $P(\cdot)$ is the probability mass function for the discrete component and $f(\cdot)$ is the truncated density for the continuous component. This type of statistical model and other variations have been studied in detail in the econometrics literature for discrete-continuous data, see \cite{takeshi1985advanced} for a reference.

\section{Simulations}
In this section, we present the simulation results that establish the validity of our approach and compare it against other approaches.

For all the simulations, we consider a setup with three treatment groups in a simulated experiment. All groups have different eventual conversion rates and a delay distribution with different means.

We compare our algorithm Delay-corrected Thompson Sampler (D-TS) against four other algorithms. 
\begin{enumerate}
    \item \textbf{Random}. As the name suggests, this algorithm randomly chooses a treatment group to display with equal probability. This can be interpreted as the classic ``split-testing". 
    \item \textbf{Naive Thompson Sampler}. This algorithm only uses the observed conversions at the assignment time and ignores the possible delays. It behaves in the same way as the standard Thompson Sampler for CTR \cite{geng2020}.
    \item \textbf{Delay-corrected UCB}.
    This is a variant of the Upper Confidence Bound (UCB) algorithm proposed by \citet{vernade2017stochastic}, where the sample size is replaced with the delay-corrected sample size plus some additional adjustments \footnote{Their paper also proposed a D-KLUCB algorithm which is claimed to be better than the D-UCB, but it is complicated to implement.}. The original paper assumes a known delay distribution, but we use estimated distribution here. The estimation follows the same EM procedure as that of our D-TS algorithm.  
    \item \textbf{Full Bayesian}. This algorithm assumes that the delay distribution follows the exponential distribution and uses the Beta priors for $\theta$ and $\lambda$. Moreover, the numerical posterior is computed and consumed by a Bayesian UCB bandit. The biggest drawback with this approach is that it is extremely time-consuming to compute, taking as much as 100 times longer than the time used by the delay-corrected methods. 
\end{enumerate}

In Table \ref{tab:benchmark}, we present the benchmark results for getting one batch of assignments from different algorithms starting from the raw log data. The benchmark test was run on a 2019 model 16-inch MacBook Pro with 16 GB Ram and 2.3 GHz 8-Core i9 Intel CPU. Each algorithm is repeated 50 times. Although these results should not be taken for their face value because the algorithms are not fully optimized for production, they show that the delay-corrected algorithm with EM procedure is reasonably fast whereas the \textit{Full Bayesian} approach is too slow for any practical use.

\begin{table}[H]
\centering
    \caption{Benchmark Results (in seconds)}
    \label{tab:benchmark}
\begin{tabular}[t]{lrrrr}
\toprule
Algorithm & min & mean & median & max\\
\midrule
Random & 0.014 & 0.018 & 0.017 & 0.038\\
Naive TS & 0.018 & 0.025 & 0.023 & 0.039\\
D-UCB & 0.345 & 0.423 & 0.402 & 0.714\\
Full Bayesian & 38.803 & 46.299 & 44.027 & 56.261\\
D-TS & 0.332 & 0.433 & 0.434 & 0.642\\
\bottomrule
\end{tabular}
\end{table}
The main metric we use to compare algorithms is cumulative regret. 
For each treatment group $k$ at the time $t$, we consider the rewards $r_t(k)$ as the total number of eventual conversions. Regrets at each time $t$ are defined as the difference between the best possible rewards at time $t$ and the rewards from the algorithm assignment plan. Mathematically, the cumulative regret is:
$$R_t = \sum_{s=1}^{t} max_k r_s(k)-r_s(k^{*})$$.

If a bandit algorithm is able to find the best group, it means that the cumulative regret should level off after some time. 

\begin{figure}
     \centering
     \begin{subfigure}[b]{0.49\linewidth}
    \centering
    \includegraphics[width=\linewidth]{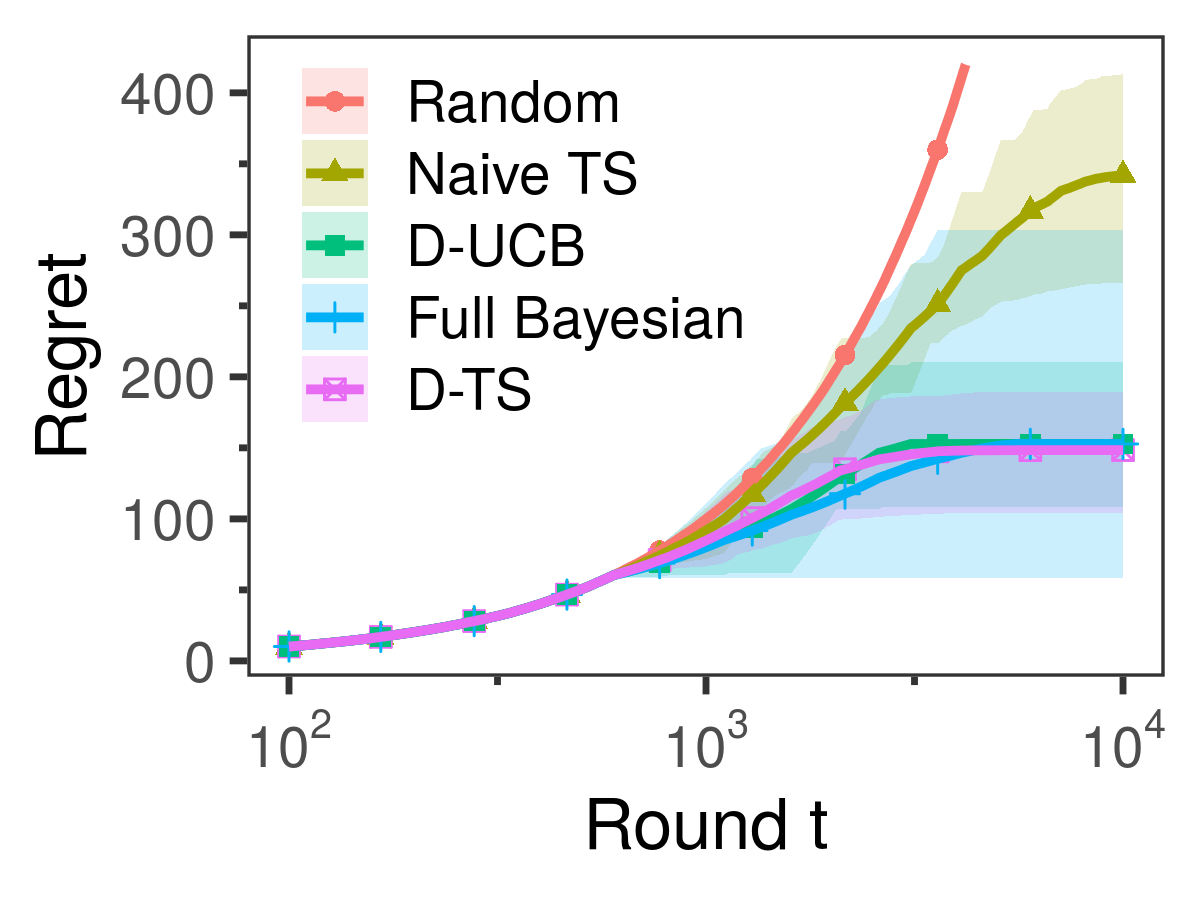}
    \caption{High CVR}
    \label{fig:simu_high_cvr}
     \end{subfigure}
     \hfill
     \begin{subfigure}[b]{0.49\linewidth}
    \centering
    \includegraphics[width=\linewidth]{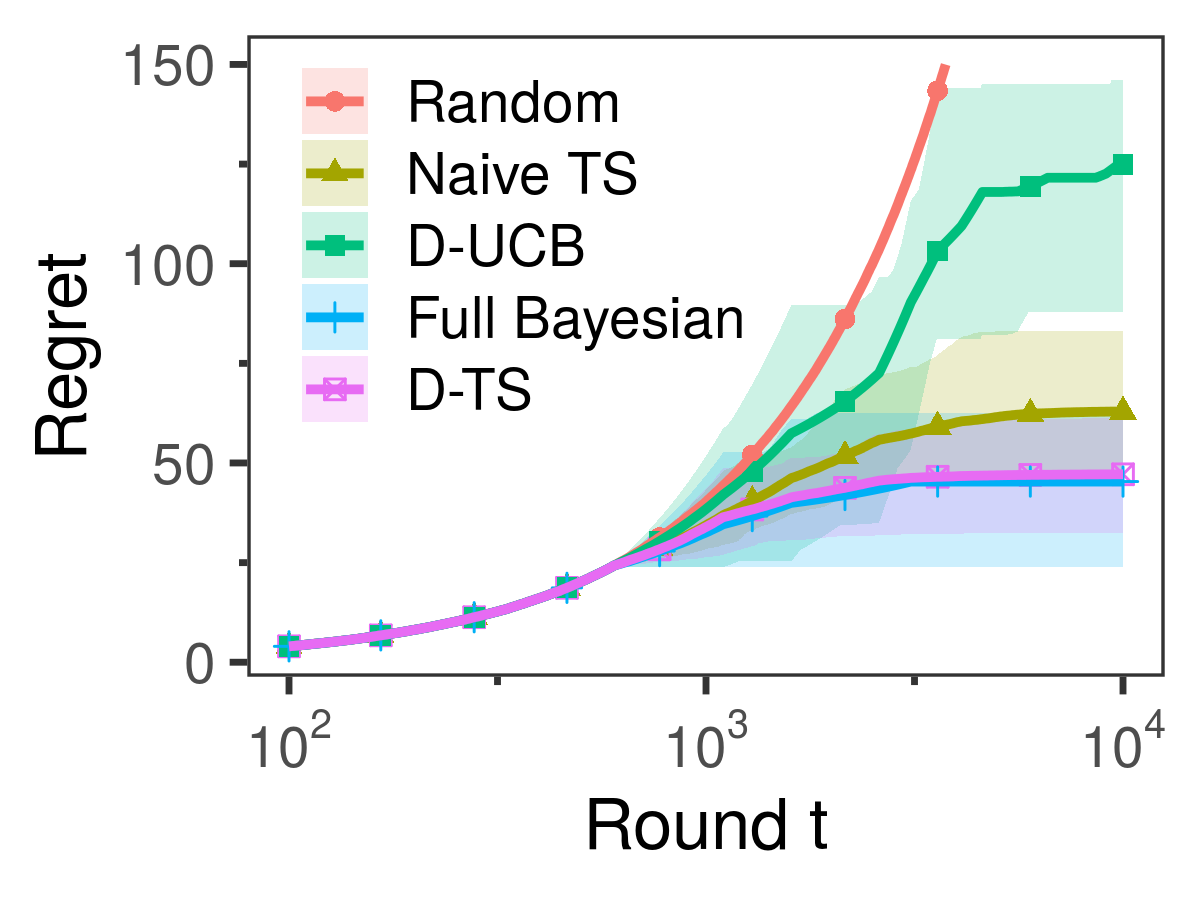}
    \caption{Low CVR}
    \label{fig:simu_low_cvr}
     \end{subfigure}
    
    \vskip\baselineskip
     \begin{subfigure}[b]{0.49\linewidth}
    \centering
    \includegraphics[width=\linewidth]{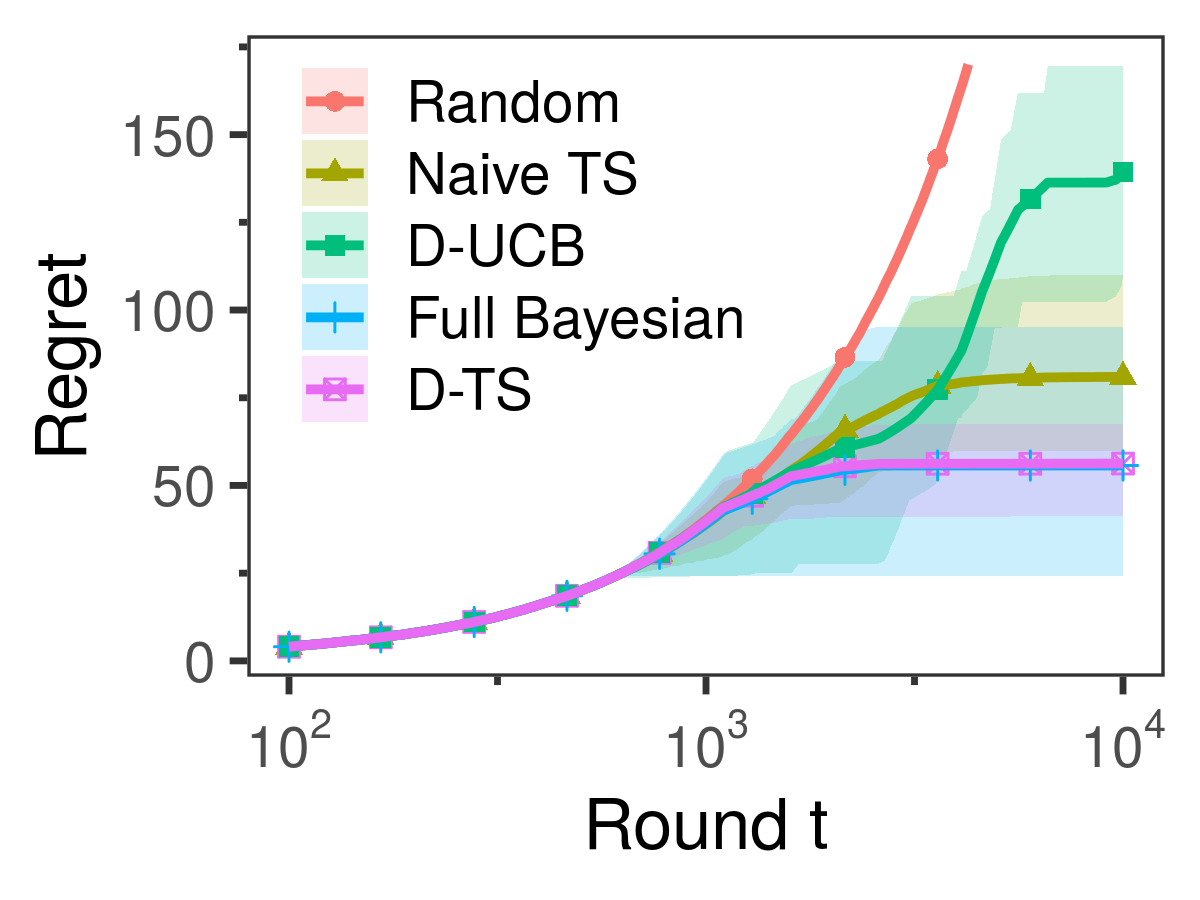}
    \caption{Weibull Distribution}
    \label{fig:simu_weibull}
     \end{subfigure}
    \hfill
   \begin{subfigure}[b]{0.49\linewidth}
    \centering
    \includegraphics[width=\linewidth]{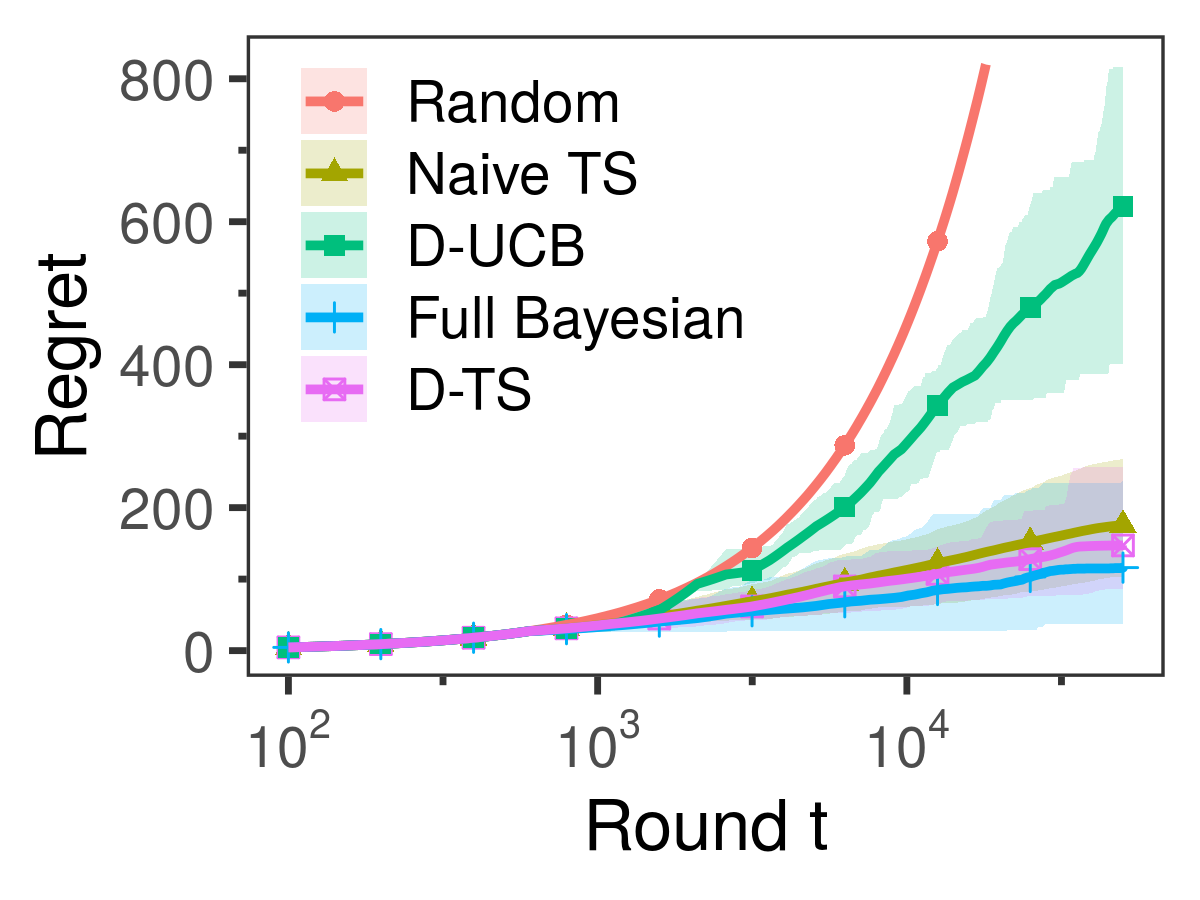}
    \caption{Criteo data}
    \label{fig:simu_criteo}
     \end{subfigure}
     \caption{Average Regrets of Various Algorithms in 4 Scenarios}
        \label{fig:simulation_results}
    \begin{flushleft}
        {\small Note:  For all experiments, there are 3 groups with different CVRs and delay distributions. The results in each experiment are averaged over 50 runs and the ribbon represents the $20_{th}$ and $80_{th}$ quantile.}
    \end{flushleft}
\end{figure}

The simulation results for 4 different environments are presented in Figure \ref{fig:simulation_results}. In Figure \ref{fig:simu_high_cvr}, we compare the cumulative
regret of the five bandit policies in a setting with relative high CVRs, $\theta_H=(0.5,0.4,0.3)$, and exponentially distributed delays with $\lambda= (1/1000, 1/750,1/500)$. In this setting, \textit{D-UCB}, \textit{Full Bayesian} and our method \textit{D-TS} are all performing well, but the \textit{Naive TS} approach takes much longer to converge. In the the low CVR setting, where $\theta_L=(0.1,0.05,0.03)$, the \textit{D-UCB} approach starts to struggle, whereas \textit{Full Bayesian} and \textit{D-TS} continue to deliver the best performance.

In Figure \ref{fig:simu_weibull}, we keep the low CVRs but change the real delay distribution to a Weibull distribution with shape parameter $k=1.5$
and the scale parameter equals the same $\lambda$ as in the previous settings. Even though our \textit{D-TS} and \textit{Full Bayesian} approaches still assume the delay distribution to be exponential, their results are not very different from the Figure \ref{fig:simu_low_cvr}, except it takes them slightly longer to converge.

In Figure \ref{fig:simu_criteo}, we generate a synthetic data using the Criteo Conversion Log data\footnote{See more details at https://labs.criteo.com/2013/12/conversion-logs-dataset/} shared in \citet{chapelle2014}. The click timestamp and delay duration for each conversion are drawn from their data. To generate variations in CVRs and delay distributions across the 3 treatment groups, we randomly drop conversions and artificially extend the delay duration for some groups. As a result, the average eventual CVRs are approximately $(0.225, 0.18, 0.135)$ and the average delays are around $(7.4, 5.6, 3.7)$ days. In this setting, the \textit{Full Bayesian} algorithm is performing the best and then followed by the \textit{D-TS} and \textit{Naive TS}. This result shows that exponentially distributed delay could be a reasonable assumption to use in practice.

\section{Deployment}
We deployed the new algorithm described in this paper online to extend JD.com's experimentation product (for more details on the experimentation platform, see \cite{geng2021comparison}). Advertisers are able to create test ad campaigns on the platform and upload multiple creatives for the algorithm to select the best one based on the CVR. If the best-performing creative attains an assignment probability larger than 90\% continuously for 24 hours, it will be declared the winner.

After the advertisers launch a creative experiment, for example, all the creatives are randomly displayed to the users in the beginning in order to collect the initial data. The system collects data on the clicks and orders whenever a creative is served. After the initial period (marked by impression counts), a service running the EM algorithm will update the CVR estimates for all the creatives in every 30 minutes. Then based on the estimated CVRs, a Thompson Sampler calculates and stores the posterior probability of being the best creatives for each creative. After that, whenever a user arrives at the e-commerce site, and the ad in the experiment is retrieved, the creative for display is chosen according to the determined probability. As more data are collected, the creative with the highest conversion rate will gradually have more chance to be displayed. 
Throughout the experiment, all the reports and relevant statistical results are displayed in a dashboard in real-time and readily available to the advertisers.

We discuss a case study based on the results from the first CVR experiment run by a large cellphone manufacturer after we launched the product. The advertiser sets up 2 creatives for the same item: one dark version and one light version. We keep track of the orders of each click for 15 days. The experiment lasted about 3 weeks, with 130 orders recorded for the dark version and 237 orders for the light version. 

\begin{figure}
    \centering
    \includegraphics[width=\linewidth]{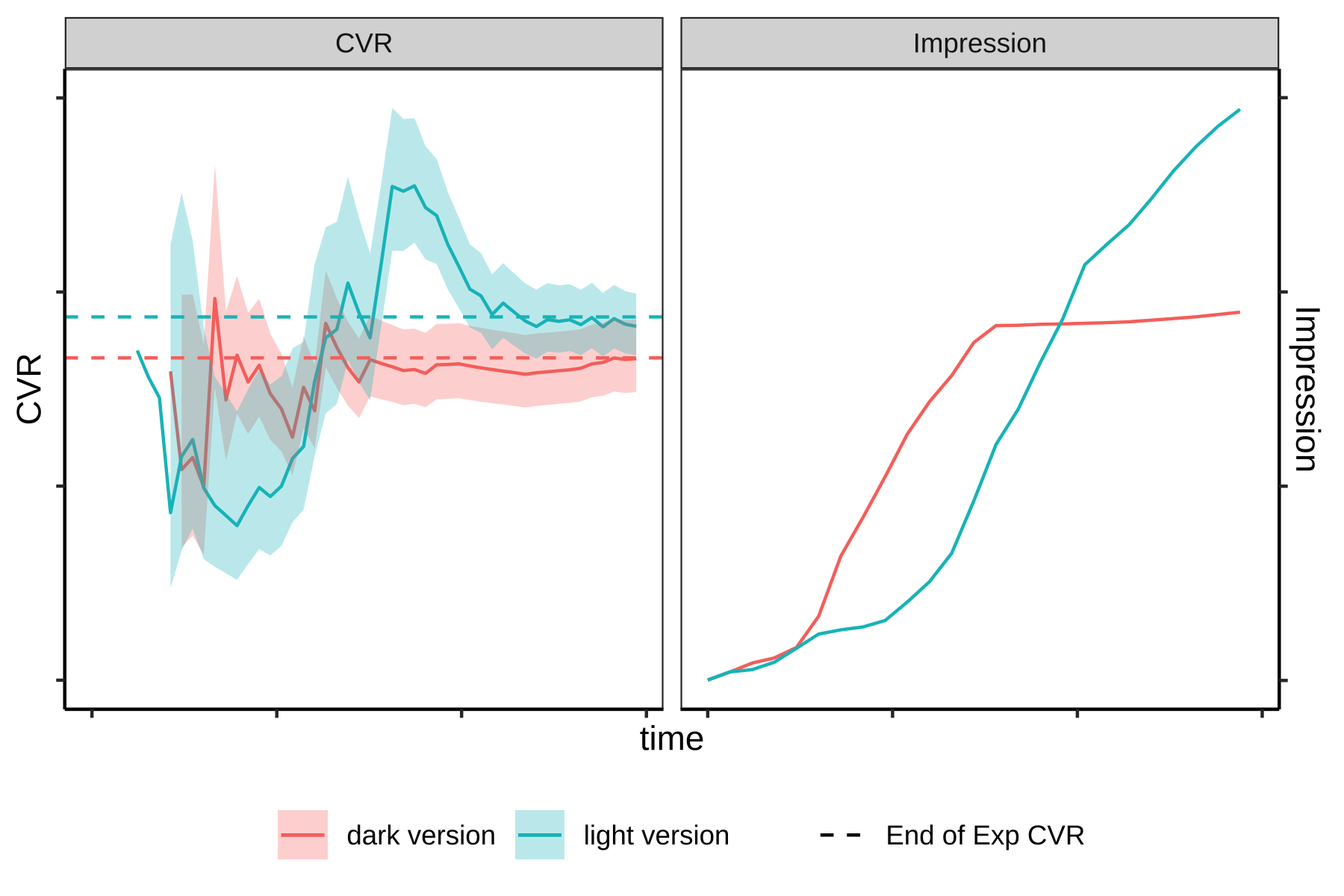}
    \caption{Case Study: An Online Creative Experiment Results}
    \begin{flushleft}
        {\small Note: Both axes start at 0. The tick values are intentionally omitted to protect business interests.}
    \end{flushleft}
    \label{fig:online-impression}
\end{figure}

In the left panel of Figure \ref{fig:online-impression}, we present the estimated delay-corrected CVRs of both versions of the creatives. The solid lines indicate the point estimate of the CVRs, whereas the dashed lines indicate the CVRs fifteen days after the experiment ends. The ribbons represent the 10th and 90th percentiles of the posterior of estimated CVRs. This left figure shows that after a period of learning, the estimated CVRs from our algorithm are able to ``predict'' the eventual CVRs of each creative after the experiment. The right panel presents the impression count for each version of the creative through the experiment. It shows the exploration and exploitation of the bandit algorithm and the fact that the algorithm eventually allocates more traffic to the higher CVR creative. 

\begin{figure}
    \centering
    \includegraphics[width=\linewidth]{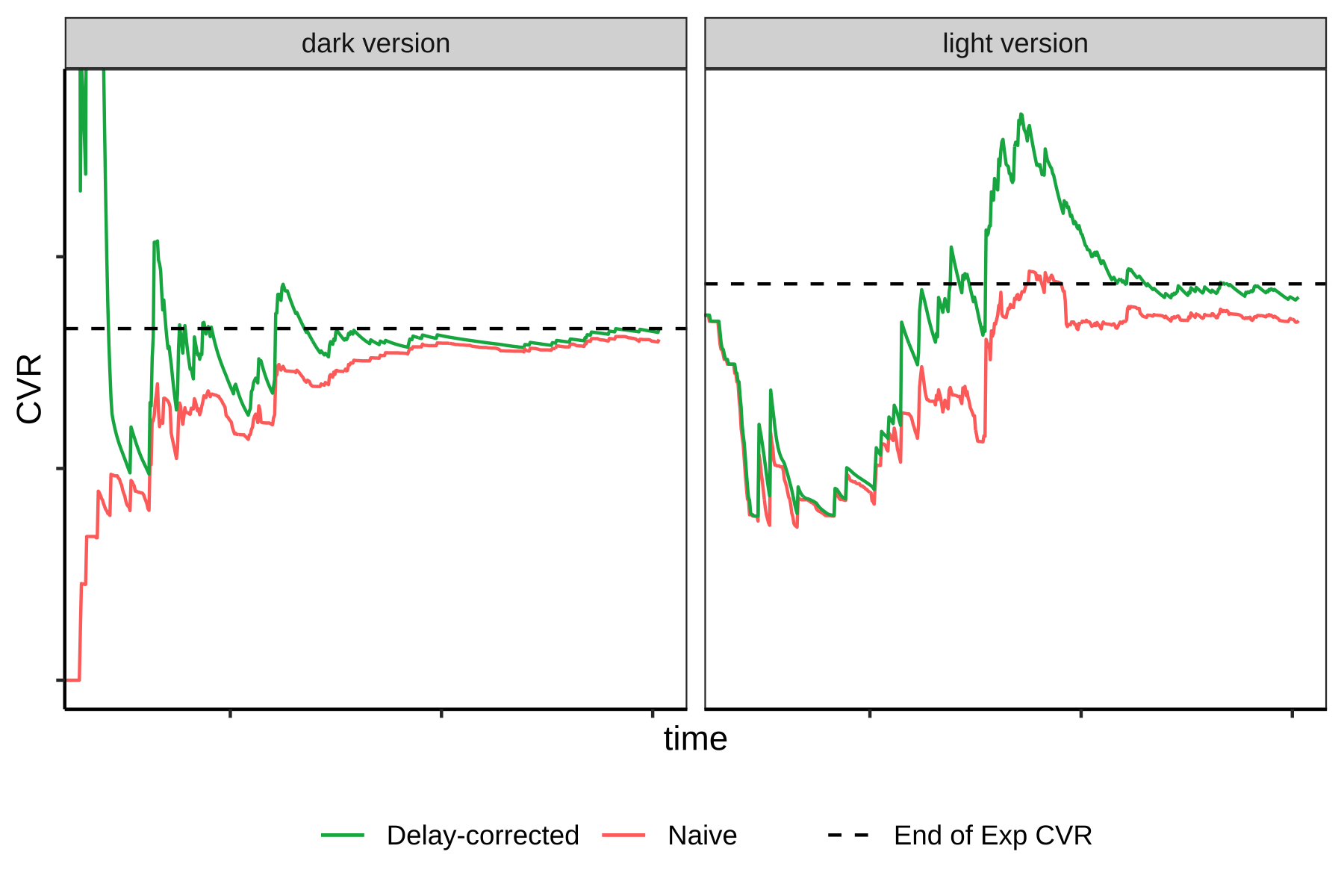}
    \caption{Delay-corrected CVR Estimate vs. Naive CVR Estimate}
    \begin{flushleft}
        {\small Note: Both axes start at 0. The tick values are intentionally omitted to protect business interests.}
    \end{flushleft}
    \label{fig:online-cvr}
\end{figure}

Although the online case study was not designed for comparing our algorithm against the other approaches,  we can still use its data to check whether our estimated CVR is a better signal for the eventual CVR compared to the naive CVR. Similar to Figure \ref{fig:cvr_comparison}, we compare the delay corrected CVR estimate against the uncorrected (or naive) CVR estimate during the experiment in Figure \ref{fig:online-cvr}. The green and red lines represent the delay-corrected CVR and naive CVR estimations, respectively, and the dashed line indicates the eventual CVR fifteen days after the experiment ends. This figure shows that our delayed-corrected estimates are much closer to the eventual CVR compared to the naive estimator and its performance improves as time progresses. In the right panel, the naive estimate greatly underestimates the eventual CVR even at the end of the experiment. The main reason is the naive estimator considers clicks that have not converted as a negative immediately, while that the delay-corrected takes into account the potential delay to conversion. This result is consistent with our argument made in Section \ref{sec:setup}.

\section{Conclusion}

An adaptive experimentation algorithm to identify the best treatment group from a set of competing treatment groups with respect to a delayed binary feedback objective was presented. This algorithm is applicable to a variety of situations common in digital advertising and has the potential to be extended to support more metrics. For our application, the algorithm powers a product that allows advertisers to identify the best creative for an ad from a set of advertising creatives for a delayed feedback outcome, i.e. conversion rate (CVR). Moreover, simulations were presented to demonstrate that the algorithm outperforms benchmarks. In addition, we discussed the deployment and presented a case study where the algorithm was used by an advertiser (a large cellphone manufacturer) to identify the optimal advertising creative for their advertising campaign. This algorithm is currently deployed in the online experimentation platform of JD.com, a large e-commerce company and a publisher of digital ads.

\bibliographystyle{ACM-Reference-Format}
\bibliography{main}


\begin{thebibliography}{23}


\ifx \showCODEN    \undefined \def \showCODEN     #1{\unskip}     \fi
\ifx \showDOI      \undefined \def \showDOI       #1{#1}\fi
\ifx \showISBNx    \undefined \def \showISBNx     #1{\unskip}     \fi
\ifx \showISBNxiii \undefined \def \showISBNxiii  #1{\unskip}     \fi
\ifx \showISSN     \undefined \def \showISSN      #1{\unskip}     \fi
\ifx \showLCCN     \undefined \def \showLCCN      #1{\unskip}     \fi
\ifx \shownote     \undefined \def \shownote      #1{#1}          \fi
\ifx \showarticletitle \undefined \def \showarticletitle #1{#1}   \fi
\ifx \showURL      \undefined \def \showURL       {\relax}        \fi
\providecommand\bibfield[2]{#2}
\providecommand\bibinfo[2]{#2}
\providecommand\natexlab[1]{#1}
\providecommand\showeprint[2][]{arXiv:#2}

\bibitem[\protect\citeauthoryear{Agarwal, Chen, and Elango}{Agarwal
  et~al\mbox{.}}{2009}]%
        {agarwal2009explore}
\bibfield{author}{\bibinfo{person}{Deepak Agarwal}, \bibinfo{person}{Bee-Chung
  Chen}, {and} \bibinfo{person}{Pradheep Elango}.}
  \bibinfo{year}{2009}\natexlab{}.
\newblock \showarticletitle{Explore/exploit schemes for web content
  optimization}. In \bibinfo{booktitle}{\emph{2009 Ninth IEEE International
  Conference on Data Mining}}. IEEE, \bibinfo{pages}{1--10}.
\newblock


\bibitem[\protect\citeauthoryear{Amemiya}{Amemiya}{1985}]%
        {takeshi1985advanced}
\bibfield{author}{\bibinfo{person}{Takeshi Amemiya}.}
  \bibinfo{year}{1985}\natexlab{}.
\newblock \bibinfo{booktitle}{\emph{Advanced econometrics}}.
\newblock \bibinfo{publisher}{Harvard university press}.
\newblock


\bibitem[\protect\citeauthoryear{Chapelle}{Chapelle}{2014}]%
        {chapelle2014}
\bibfield{author}{\bibinfo{person}{Olivier Chapelle}.}
  \bibinfo{year}{2014}\natexlab{}.
\newblock \showarticletitle{Modeling delayed feedback in display advertising}.
  In \bibinfo{booktitle}{\emph{Proceedings of the 20th ACM SIGKDD international
  conference on Knowledge discovery and data mining}}.
  \bibinfo{pages}{1097--1105}.
\newblock


\bibitem[\protect\citeauthoryear{Chapelle and Li}{Chapelle and Li}{2011}]%
        {chapelle2011empirical}
\bibfield{author}{\bibinfo{person}{Olivier Chapelle} {and}
  \bibinfo{person}{Lihong Li}.} \bibinfo{year}{2011}\natexlab{}.
\newblock \showarticletitle{An empirical evaluation of thompson sampling}.
\newblock \bibinfo{journal}{\emph{Advances in neural information processing
  systems}}  \bibinfo{volume}{24} (\bibinfo{year}{2011}),
  \bibinfo{pages}{2249--2257}.
\newblock


\bibitem[\protect\citeauthoryear{Dupret and Lalmas}{Dupret and Lalmas}{2013}]%
        {dupret2013absence}
\bibfield{author}{\bibinfo{person}{Georges Dupret} {and}
  \bibinfo{person}{Mounia Lalmas}.} \bibinfo{year}{2013}\natexlab{}.
\newblock \showarticletitle{Absence time and user engagement: evaluating
  ranking functions}. In \bibinfo{booktitle}{\emph{Proceedings of the sixth ACM
  international conference on Web search and data mining}}.
  \bibinfo{pages}{173--182}.
\newblock


\bibitem[\protect\citeauthoryear{Geng, Lin, and Nair}{Geng
  et~al\mbox{.}}{2020}]%
        {geng2020}
\bibfield{author}{\bibinfo{person}{Tong Geng}, \bibinfo{person}{Xiliang Lin},
  {and} \bibinfo{person}{Harikesh~S Nair}.} \bibinfo{year}{2020}\natexlab{}.
\newblock \showarticletitle{Online Evaluation of Audiences for Targeted
  Advertising via Bandit Experiments}. In \bibinfo{booktitle}{\emph{Proceedings
  of the AAAI Conference on Artificial Intelligence}},
  Vol.~\bibinfo{volume}{34}. \bibinfo{pages}{13273--13279}.
\newblock


\bibitem[\protect\citeauthoryear{Geng, Lin, Nair, Hao, Xiang, and Fan}{Geng
  et~al\mbox{.}}{2021}]%
        {geng2021comparison}
\bibfield{author}{\bibinfo{person}{Tong Geng}, \bibinfo{person}{Xiliang Lin},
  \bibinfo{person}{Harikesh~S Nair}, \bibinfo{person}{Jun Hao},
  \bibinfo{person}{Bin Xiang}, {and} \bibinfo{person}{Shurui Fan}.}
  \bibinfo{year}{2021}\natexlab{}.
\newblock \showarticletitle{Comparison Lift: Bandit-based Experimentation
  System for Online Advertising}. In \bibinfo{booktitle}{\emph{Proceedings of
  the AAAI Conference on Artificial Intelligence}}, Vol.~\bibinfo{volume}{35}.
  \bibinfo{pages}{15117--15126}.
\newblock


\bibitem[\protect\citeauthoryear{Hunter and Lange}{Hunter and Lange}{2004}]%
        {hunter2004tutorial}
\bibfield{author}{\bibinfo{person}{David~R Hunter} {and}
  \bibinfo{person}{Kenneth Lange}.} \bibinfo{year}{2004}\natexlab{}.
\newblock \showarticletitle{A tutorial on MM algorithms}.
\newblock \bibinfo{journal}{\emph{The American Statistician}}
  \bibinfo{volume}{58}, \bibinfo{number}{1} (\bibinfo{year}{2004}),
  \bibinfo{pages}{30--37}.
\newblock


\bibitem[\protect\citeauthoryear{Johari, Pekelis, and Walsh}{Johari
  et~al\mbox{.}}{2015}]%
        {johari2015always}
\bibfield{author}{\bibinfo{person}{Ramesh Johari}, \bibinfo{person}{Leo
  Pekelis}, {and} \bibinfo{person}{David~J Walsh}.}
  \bibinfo{year}{2015}\natexlab{}.
\newblock \showarticletitle{Always valid inference: Bringing sequential
  analysis to A/B testing}.
\newblock \bibinfo{journal}{\emph{arXiv preprint arXiv:1512.04922}}
  (\bibinfo{year}{2015}).
\newblock


\bibitem[\protect\citeauthoryear{Joulani, Gyorgy, and Szepesv{\'a}ri}{Joulani
  et~al\mbox{.}}{2013}]%
        {joulani2013online}
\bibfield{author}{\bibinfo{person}{Pooria Joulani}, \bibinfo{person}{Andras
  Gyorgy}, {and} \bibinfo{person}{Csaba Szepesv{\'a}ri}.}
  \bibinfo{year}{2013}\natexlab{}.
\newblock \showarticletitle{Online learning under delayed feedback}. In
  \bibinfo{booktitle}{\emph{International Conference on Machine Learning}}.
  PMLR, \bibinfo{pages}{1453--1461}.
\newblock


\bibitem[\protect\citeauthoryear{Kalbfleisch and Prentice}{Kalbfleisch and
  Prentice}{2011}]%
        {prentice2011}
\bibfield{author}{\bibinfo{person}{John~D Kalbfleisch} {and}
  \bibinfo{person}{Ross~L Prentice}.} \bibinfo{year}{2011}\natexlab{}.
\newblock \bibinfo{booktitle}{\emph{The statistical analysis of failure time
  data}}.
\newblock \bibinfo{publisher}{John Wiley \& Sons}.
\newblock


\bibitem[\protect\citeauthoryear{Kohavi, Tang, and Xu}{Kohavi
  et~al\mbox{.}}{2020}]%
        {kohavi2020}
\bibfield{author}{\bibinfo{person}{Ron Kohavi}, \bibinfo{person}{Diane Tang},
  {and} \bibinfo{person}{Ya Xu}.} \bibinfo{year}{2020}\natexlab{}.
\newblock \bibinfo{booktitle}{\emph{Trustworthy online controlled experiments:
  A practical guide to a/b testing}}.
\newblock \bibinfo{publisher}{Cambridge University Press}.
\newblock


\bibitem[\protect\citeauthoryear{Lattimore and Szepesv{\'a}ri}{Lattimore and
  Szepesv{\'a}ri}{2020}]%
        {lattimore2020}
\bibfield{author}{\bibinfo{person}{Tor Lattimore} {and} \bibinfo{person}{Csaba
  Szepesv{\'a}ri}.} \bibinfo{year}{2020}\natexlab{}.
\newblock \bibinfo{booktitle}{\emph{Bandit algorithms}}.
\newblock \bibinfo{publisher}{Cambridge University Press}.
\newblock


\bibitem[\protect\citeauthoryear{Lehmann, Lalmas, Yom-Tov, and Dupret}{Lehmann
  et~al\mbox{.}}{2012}]%
        {lehmann2012models}
\bibfield{author}{\bibinfo{person}{Janette Lehmann}, \bibinfo{person}{Mounia
  Lalmas}, \bibinfo{person}{Elad Yom-Tov}, {and} \bibinfo{person}{Georges
  Dupret}.} \bibinfo{year}{2012}\natexlab{}.
\newblock \showarticletitle{Models of user engagement}. In
  \bibinfo{booktitle}{\emph{International conference on user modeling,
  adaptation, and personalization}}. Springer, \bibinfo{pages}{164--175}.
\newblock


\bibitem[\protect\citeauthoryear{Li, Chu, Langford, and Schapire}{Li
  et~al\mbox{.}}{2010}]%
        {li2010contextual}
\bibfield{author}{\bibinfo{person}{Lihong Li}, \bibinfo{person}{Wei Chu},
  \bibinfo{person}{John Langford}, {and} \bibinfo{person}{Robert~E Schapire}.}
  \bibinfo{year}{2010}\natexlab{}.
\newblock \showarticletitle{A contextual-bandit approach to personalized news
  article recommendation}. In \bibinfo{booktitle}{\emph{Proceedings of the 19th
  international conference on World wide web}}. \bibinfo{pages}{661--670}.
\newblock


\bibitem[\protect\citeauthoryear{Liu, Downe, and Reid}{Liu
  et~al\mbox{.}}{2019}]%
        {liu2019multi}
\bibfield{author}{\bibinfo{person}{Larkin Liu}, \bibinfo{person}{Richard
  Downe}, {and} \bibinfo{person}{Joshua Reid}.}
  \bibinfo{year}{2019}\natexlab{}.
\newblock \showarticletitle{Multi-armed bandit strategies for non-stationary
  reward distributions and delayed feedback processes}.
\newblock \bibinfo{journal}{\emph{arXiv preprint arXiv:1902.08593}}
  (\bibinfo{year}{2019}).
\newblock


\bibitem[\protect\citeauthoryear{Pike-Burke, Agrawal, Szepesvari, and
  Grunewalder}{Pike-Burke et~al\mbox{.}}{2018}]%
        {pike2018bandits}
\bibfield{author}{\bibinfo{person}{Ciara Pike-Burke}, \bibinfo{person}{Shipra
  Agrawal}, \bibinfo{person}{Csaba Szepesvari}, {and} \bibinfo{person}{Steffen
  Grunewalder}.} \bibinfo{year}{2018}\natexlab{}.
\newblock \showarticletitle{Bandits with delayed, aggregated anonymous
  feedback}. In \bibinfo{booktitle}{\emph{International Conference on Machine
  Learning}}. PMLR, \bibinfo{pages}{4105--4113}.
\newblock


\bibitem[\protect\citeauthoryear{Schwartz, Bradlow, and Fader}{Schwartz
  et~al\mbox{.}}{2017}]%
        {schwartz2017customer}
\bibfield{author}{\bibinfo{person}{Eric~M Schwartz}, \bibinfo{person}{Eric~T
  Bradlow}, {and} \bibinfo{person}{Peter~S Fader}.}
  \bibinfo{year}{2017}\natexlab{}.
\newblock \showarticletitle{Customer acquisition via display advertising using
  multi-armed bandit experiments}.
\newblock \bibinfo{journal}{\emph{Marketing Science}} \bibinfo{volume}{36},
  \bibinfo{number}{4} (\bibinfo{year}{2017}), \bibinfo{pages}{500--522}.
\newblock


\bibitem[\protect\citeauthoryear{Scott}{Scott}{2010}]%
        {scott2010}
\bibfield{author}{\bibinfo{person}{Steven~L Scott}.}
  \bibinfo{year}{2010}\natexlab{}.
\newblock \showarticletitle{A modern Bayesian look at the multi-armed bandit}.
\newblock \bibinfo{journal}{\emph{Applied Stochastic Models in Business and
  Industry}} \bibinfo{volume}{26}, \bibinfo{number}{6} (\bibinfo{year}{2010}),
  \bibinfo{pages}{639--658}.
\newblock


\bibitem[\protect\citeauthoryear{Scott}{Scott}{2015}]%
        {scott2015multi}
\bibfield{author}{\bibinfo{person}{Steven~L Scott}.}
  \bibinfo{year}{2015}\natexlab{}.
\newblock \showarticletitle{Multi-armed bandit experiments in the online
  service economy}.
\newblock \bibinfo{journal}{\emph{Applied Stochastic Models in Business and
  Industry}} \bibinfo{volume}{31}, \bibinfo{number}{1} (\bibinfo{year}{2015}),
  \bibinfo{pages}{37--45}.
\newblock


\bibitem[\protect\citeauthoryear{Vernade, Capp{\'e}, and Perchet}{Vernade
  et~al\mbox{.}}{2017}]%
        {vernade2017stochastic}
\bibfield{author}{\bibinfo{person}{Claire Vernade}, \bibinfo{person}{Olivier
  Capp{\'e}}, {and} \bibinfo{person}{Vianney Perchet}.}
  \bibinfo{year}{2017}\natexlab{}.
\newblock \showarticletitle{Stochastic bandit models for delayed conversions}.
\newblock \bibinfo{journal}{\emph{arXiv preprint arXiv:1706.09186}}
  (\bibinfo{year}{2017}).
\newblock


\bibitem[\protect\citeauthoryear{Vernade, Carpentier, Lattimore, Zappella,
  Ermis, and Brueckner}{Vernade et~al\mbox{.}}{2020}]%
        {vernade2020linear}
\bibfield{author}{\bibinfo{person}{Claire Vernade}, \bibinfo{person}{Alexandra
  Carpentier}, \bibinfo{person}{Tor Lattimore}, \bibinfo{person}{Giovanni
  Zappella}, \bibinfo{person}{Beyza Ermis}, {and} \bibinfo{person}{Michael
  Brueckner}.} \bibinfo{year}{2020}\natexlab{}.
\newblock \showarticletitle{Linear bandits with stochastic delayed feedback}.
  In \bibinfo{booktitle}{\emph{International Conference on Machine Learning}}.
  PMLR, \bibinfo{pages}{9712--9721}.
\newblock


\bibitem[\protect\citeauthoryear{Zhou, Xu, and Blanchet}{Zhou
  et~al\mbox{.}}{2019}]%
        {zhou2019learning}
\bibfield{author}{\bibinfo{person}{Zhengyuan Zhou}, \bibinfo{person}{Renyuan
  Xu}, {and} \bibinfo{person}{Jose Blanchet}.} \bibinfo{year}{2019}\natexlab{}.
\newblock \showarticletitle{Learning in generalized linear contextual bandits
  with stochastic delays}.
\newblock \bibinfo{journal}{\emph{Advances in Neural Information Processing
  Systems}}  \bibinfo{volume}{32} (\bibinfo{year}{2019}),
  \bibinfo{pages}{5197--5208}.
\newblock


\end{thebibliography}

\appendix

\section{Appendix}

\subsection{Unbiasedness of delay-corrected estimator}\label{appendix:unbiasedness}
\begin{proposition}
If the delay variable $D_i$ is independent and identically distributed across $i$, then   
$$
    \widehat{\theta_t}=\frac{N_t^{convert}}{\sum_{s=1}^{t}n_sP(D \leq t-s)}
$$ is an unbiased estimator for $\theta = E[C_i]$.
\end{proposition}
\begin{proof}
The $\widehat{\theta_t}$ can be re-written as
$$    
\widehat{\theta_t}=\frac{\sum_{s=1}^{t} \sum_{i^s=1}^{n_s} C_{i^s} 1\left\{D_{i^s} \leq t-s\right\}}{\sum_{s=1}^{t}n_sP(D \leq t-s)}.
$$
So
\begin{align*}
    E[\widehat{\theta_t}] & = \frac{E[\sum_{s=1}^{t} \sum_{i^s=1}^{n_s} C_{i^s} 1\left\{D_{i^s} \leq t-s\right\}]}{\sum_{s=1}^{t}n_sP(D \leq t-s)} \\
    &= \frac{\sum_{s=1}^{t} \sum_{i^s=1}^{n_s} E[C_{i^s} ]E[1\left\{D_{i^s} \leq t-s\right\}]}{\sum_{s=1}^{t}n_sP(D \leq t-s)} \\
    &= \frac{\sum_{s=1}^{t} \sum_{i=1}^{n_s} \theta P(D_{i^s} \leq t-s)}{\sum_{s=1}^{t}n_sP(D \leq t-s)}\\
    &= \theta\frac{\sum_{s=1}^{t} n_s  P(D \leq t-s) }{\sum_{s=1}^{t}n_sP(D \leq t-s)}\\
    &=\theta 
    \end{align*}
\end{proof}

\section{Online Resources}
The R source code for the simulation exercises will be available upon publication.
\end{document}